\documentclass[aps,preprint,amsmath,showpacs,nofootinbib,tightenlines]{revtex4-1}

\usepackage{graphicx}  %
\usepackage{amsmath}
\usepackage{bm}  %
\usepackage{ulem}  %

\newcommand{\bea}{\begin{eqnarray}}
\newcommand{\eea}{\end{eqnarray}}
\newcommand{\beq}{\begin{equation}}
\newcommand{\eeq}{\end{equation}}
\newcommand{\bqa}{\begin{eqnarray}}
\newcommand{\eqa}{\end{eqnarray}}

\def\mqo2{{\!\!\!}}

\renewcommand{\emph}[1]{{\it #1}}

\newcommand{\psid}{\psi^\dagger}
\newcommand{\phid}{\phi^\dagger}
\newcommand{\bra}[1]{\langle #1 |}

\newcommand{\Lag}{\mathcal{L}}

\newcommand{\rv}{\bm{r}}

\newcommand{\eff}{\mathrm{eff}}

\newcommand{\Tr}{\mathrm{Tr}}
\newcommand{\ket}[1]{|#1\rangle}

\newcommand{\deep}{\mathrm{deep}}
\newcommand{\full}{\mathrm{full}}

\begin{document}

\title{
Lindblad Equation for the Inelastic Loss of Ultracold Atoms}
\author{Eric Braaten}
\affiliation{Department of Physics,
         The Ohio State University, Columbus, OH\ 43210, USA}
\author{H.-W. Hammer}
\affiliation{Institut f\"ur Kernphysik, Technische Universit\"at Darmstadt,
64289\ Darmstadt, Germany}
\affiliation{ExtreMe Matter Institute EMMI, GSI Helmholtzzentrum f\"ur 
Schwerionenforschung, 64291\ Darmstadt, Germany}
\author{G.~Peter Lepage}
\affiliation{Laboratory of Elementary Particle Physics, Cornell University,
Ithaca, New York\ 14583, USA}
\date{\today}

\begin{abstract}
The loss of ultracold trapped atoms due to deeply inelastic reactions has
previously been taken into account in effective field
theories for low-energy atoms
by adding local anti-Hermitian terms to the effective Hamiltonian.
Here we show that when multi-atom systems are considered, an additional 
modification is required in the equation governing the density matrix. 
We define an effective density matrix by tracing over the states
containing high-momentum atoms produced by deeply inelastic reactions.
We show that it satisfies a Lindblad equation, with local Lindblad operators
determined by the local anti-Hermitian terms in the effective Hamiltonian.
We use the Lindblad equation to derive the universal relation for the two-atom
inelastic loss rate for fermions with two spin states and  the universal
relation for the three-atom inelastic loss rate for identical bosons.
\end{abstract}

\smallskip
\maketitle

\section{Introduction}
\label{sec:intro}

The development of technologies to trap and cool neutral atoms has led to 
the emergence of cold atom physics as a new subfield of atomic physics.
The atoms can be cooled to temperatures that are orders of magnitude smaller 
than the tiny differences between the hyperfine energy levels of the atoms.
Many of the most important loss processes for ultracold atoms involve deeply inelastic reactions,
which produce atoms with kinetic energies much larger than those of the cold trapped atoms.
One such process is three-body recombination,
in which a collision of three low-energy atoms results in the binding of two of the atoms
into a diatomic molecule.
If the binding energy of the molecule is large compared to the energy scale of the cold atoms,
the momenta of the molecule and the recoiling atom
are often large enough for them to escape from the trapping potential for the atoms.

Ultracold atoms can be described by a local nonrelativistic effective field theory 
for which the coupling constant is  the scattering length \cite{Braaten:2004rn}.
Local effective field theories can be applied most rigorously to a system in which there is an energy gap
separating the low-energy particles described explicitly by the effective field theory
from the higher-momentum particles \cite{Georgi:1994qn,Kaplan:1995uv,Manohar:1996cq,Shankar:1996vk,Burgess:2007pt}.  
In a system consisting of low-energy atoms,
conservation of energy ensures that a high-momentum atom
can only appear through a virtual state that, by the time-energy uncertainty principle, has a short lifetime.
During that short lifetime, the high-momentum atom can propagate only over a short distance.
Its effects on low-energy atoms are therefore restricted to short distances.
These effects can be reproduced by local Hermitian operators in the 
Hamiltonian for the effective field theory.

Local effective field theories have also been applied to systems
with deeply inelastic reactions that produce particles with momenta too large to be
described accurately within the effective field theory.
For example,  a deeply inelastic three-body recombination
event produces a molecule and an atom whose momenta 
may be outside the domain of validity of the effective theory.
The large energy release from the  inelastic reaction comes from the  conversion of rest energy 
into kinetic energy.
The standard argument for the locality of the effective field theory does not apply.
The particles with large momenta produced by the inelastic reaction can propagate over arbitrarily
long distances, so their effects on low-energy particles are not 
obviously restricted to short distances.
Nevertheless, there are general arguments based on the uncertainty principle
that indicate that their effects can be taken into account systematically through 
local anti-Hermitian operators  in the effective  Hamiltonian \cite{BHL:1607}.
The effective Hamiltonian can be expressed as  $H_{\rm eff} = H -iK$,
where $H$ and $K$ are Hermitian. 

The dynamics of a multi-atom system with deeply inelastic reactions
is conveniently described by a density matrix.
A system that starts as a pure quantum state with $n$ low-energy atoms evolves 
into a mixed quantum state
that is a superposition of a state with $n$ low-energy atoms and states with fewer low-energy atoms, 
as the inelastic reactions shift probability from the low-energy atoms into
the high-momentum atoms. An effective density matrix $\rho$
can be defined by tracing the density matrix over states containing high-momentum atoms.
Naively we might expect  the time evolution equation for $\rho$ to be
  \begin{eqnarray}
 i\hbar \frac{d\ }{dt}\rho &\overset{?}{=} &  H_\eff \rho- \rho H_\eff
=  \left[H, \rho\right] - i \left\{K,\rho\right\}.
\label{eq:drhodt-naive}
\end{eqnarray}
As we will demonstrate in this paper, the correct evolution equation
for the effective density matrix is the {\it Lindblad equation}
\cite{Lindblad:1975ef,Gorini:1975nb}, which has an additional term.
The Lindblad equation arises in the quantum information theory
of  {\it open quantum systems}.
An  open quantum system
consists of all the degrees of freedom
of both the subsystem of interest and the environment. 
Under special circumstances, the density matrix for the subsystem evolves 
in time according to the Lindblad equation.
In the Lindblad equation for the density matrix of an effective field theory
obtained by integrating out deeply  inelastic reactions,
the additional Lindblad term is local, and it
can be deduced from the local anti-Hermitian terms
in the effective Hamiltonian.

An open quantum system in which the subsystem of interest is a  field theory 
is called an {\it open effective field theory} \cite{Grozdanov:2013dba,Burgess:2014eoa}.
Grozdanov and Polonyi have proposed that an open effective field theory for the hydrodynamic modes 
of a quantum field theory can be used as a framework for deriving 
dissipative hydrodynamics \cite{Grozdanov:2013dba}.
Burgess, Holman, Tasinato, and Williams have applied open effective field theory 
to the super-Hubble modes  of primordial quantum fluctuations 
in the early universe \cite{Burgess:2014eoa,Burgess:2015ajz}.
In the stochastic inflation framework, the master equation is  the Lindblad equation.
Since the density matrix for an effective field theory in which deeply inelastic reaction products 
have been integrated out satisfies the Lindblad equation,
this system can also be regarded as an open effective field theory \cite{BHL:1607}.
In this case, the environment consists of the high-momentum particles 
produced by the deeply inelastic reactions.

The paper is organized as follows.
In Section~\ref{sec:DensityMatrix}, we summarize the basic properties of the density matrix 
and we introduce the Lindblad equation.
In Section~\ref{sec:Effective Theory}, we explain how integrating out deeply inelastic reaction
products results in local operators in the effective Lagrangian density.
We also explain why the effective density matrix obtained by tracing over states that include
deeply inelastic reaction products must satisfy the Lindblad equation.
In Section~\ref{sec:AtomLoss}, we apply the Lindblad equation to the mean 
number of low-energy atoms.
We derive the universal relation for the two-atom inelastic loss rate for fermions with two spin states 
and  the universal relation for the three-atom inelastic loss rate for identical bosons.
Our results  are summarized briefly in Section~\ref{sec:Summary}.
In an Appendix, we demonstrate how the Lindblad equation for the density matrix emerges from 
the diagrammatic analysis of a simple field theory model with a deeply
inelastic two-atom scattering reaction.

\section{Density Matrix}
\label{sec:DensityMatrix}

The density matrix can be used to describe  pure quantum states, mixed quantum states,
and statistical ensembles of quantum states.
The dynamics of a multi-atom system with atom losses
must be described by the density matrix in order to be able to track
the losses in the different $n$-atom sectors.
To lay the foundation for the rest of the paper,
we summarize the key properties of the density matrix 
and we introduce the Lindblad equation.

\subsection{General Properties}

A pure state in a quantum system
can be represented by a vector $|\psi\rangle$ in a complex vector space.
The time evolution of the quantum state is  
described by the Schr\"odinger equation:
$i \hbar (d/dt)|\psi\rangle =  H |\psi\rangle$,
where the Hamiltonian $H$ is a Hermitian operator.
The time evolution preserves the norm $\langle \psi | \psi \rangle$ of a state.

An arbitrary statistical ensemble of quantum states
can be represented by a density matrix $\rho$.
The density matrix has the following basic properties:
\begin{itemize}
\item
It is Hermitian:  $\rho^\dagger = \rho$.
\item
It is positive: $\langle \psi | \rho | \psi \rangle \ge 0$ for  all nonzero states $|\psi\rangle$.
\item
It can be normalized to have unit trace: ${\rm Tr}(\rho) = 1$.
\end{itemize}
The density matrix can also describe a pure quantum state,
in which case $\rho^2 = \rho$.

The time evolution of the density matrix is described by 
the von Neumann equation:
\begin{equation}
i \hbar \frac{d\ }{dt} \rho =  [ H, \rho],
\label{evol:Schrodinger}
\end{equation}
which follows from the Schr\"odinger equation for $|\psi\rangle$.
This evolution equation has the following properties:
\begin{itemize}
\item
It is linear in $\rho$.
\item
It preserves the trace of $\rho$.
It therefore respects the normalization condition ${\rm Tr}(\rho) = 1$.
\item
It is {\it Markovian}:    the future is determined by the present only.
There is no additional dependence on the past history.
\end{itemize}
The system average of an operator $O$
can be expressed as the trace  of its product with the density matrix:
\begin{equation}
\langle O \rangle = {\rm Tr}(\rho O ).
\label{d<O>}
\end{equation}
If the operator has no explicit time dependence,
the time evolution of the system average is determined by
the evolution equation of $\rho$ in Eq.~\eqref{evol:Schrodinger}.

\subsection{Lindblad equation}
\label{sec:Lindblad}

In the field of  quantum information theory, the full system 
is often separated into a {\it subsystem} of interest and its {\it environment}  \cite{Preskill}.
Of particular interest is the decoherence of the subsystem due to the effects of the environment.
The basis for the quantum states of the full system can be chosen to be 
direct products of  basis states for the subsystem and those for the environment.
A  density matrix $\rho$ for the subsystem can be obtained 
from the density matrix $\rho_\full$ for the full system 
by the partial trace over the states of the environment:
    \begin{equation}
    \label{eq:rho_eff-qi}
        \rho = \Tr_{\rm environment} \left(\rho_\full \right).
    \end{equation}    
The density matrix for the full system evolves according to the
von Neumann equation in Eq.~\eqref{evol:Schrodinger}.
Given the initial density matrix $ \rho(t=0)$ for the subsystem
and a specified initial state of the environment, the von Neumann equation 
can in principle be solved to determine $ \rho(t)$ at future times.
It is possible in principle to construct a self-contained differential equation
for $ \rho(t)$, but its evolution is non-Markovian  \cite{Preskill}.
The time derivative of $\rho(t)$  is determined
by the present density matrix and by its history from time 0 to $t$.
The previous history is needed to take into account the flow of information
between the subsystem and the environment.

There are situations in which the time evolution of the subsystem
can be described approximately by a self-contained Markovian differential equation.
The time during which the subsystem is observed
must be much larger than the time scale for correlations between
the subsystem and the environment.
We must also restrict our consideration to the low-frequency behavior of the subsystem,
which can be accomplished by smoothing out over times larger than the correlation time scale.
The density matrix $\rho$ for the subsystem necessarily satisfies
the three basic properties listed before Eq.~\eqref{evol:Schrodinger}:
it is Hermitian, it is positive, and it can be normalized.
It would be desirable for its time evolution to also be in accord with the three
basic properties listed after Eq.~\eqref{evol:Schrodinger}:
it should be linear in $\rho$, it should preserve the trace of $\rho$, and it should be Markovian. 
In 1976, Lindblad and, independently, Gorini, Kossakowski, and Sudarshan
showed that these conditions together with one additional technical condition require the
time evolution equation to have the form \cite{Lindblad:1975ef,Gorini:1975nb}

\begin{equation}
i \hbar \frac{d\ }{dt} \rho =
 [ H, \rho] - \frac{i}{2} \sum_n
\left( L_n^\dagger L_n \rho + \rho L_n^\dagger L_n
-2 L_n \rho L_n^\dagger \right),
\label{evol:Lindblad}
\end{equation}
where  $H$ is a Hermitian operator
and the $L_n$'s are an additional set of operators called {\it Lindblad operators}.
The evolution equation in Eq.~\eqref{evol:Lindblad} is called the {\it Lindblad equation}.
The linear operator acting on $\rho$ defined by the right side of
Eq.~\eqref{evol:Lindblad} is called the {\it Lindbladian}.
Thus the Lindbladian is the generator of the time evolution of the density matrix.

The additional technical condition required to derive the Lindblad equation is that
the linear operator that determines the time evolution of $\rho$ 
should be completely positive  \cite{Lindblad:1975ef,Gorini:1975nb}.
For any complex vector space, this linear operator 
has a natural extension to an operator acting on the tensor product
of the quantum-state spaces for the subsystem and the other complex vector space. 
Complete positivity requires the extension of the operator to be positive
in the larger space.
An accessible derivation of the Lindblad equation can be found in lecture notes 
on quantum information and computation by John Preskill \cite{Preskill}.

The Lindblad equation in Eq.~\eqref{evol:Lindblad} can be expressed in the form
\begin{equation}
i \hbar \frac{d\ }{dt} \rho =
 [ H, \rho] - i \left\{K,\rho\right\}
 + i \sum_n L_n \rho L_n^\dagger ,
\label{evol:Lindblad-2}
\end{equation}
where the Hermitian operator $K$ is
\begin{equation}
K =\frac12\sum_n L_n^\dagger L_n.
\label{eq:K-Ln}
\end{equation}
Comparison with the naive evolution equation in Eq.~\eqref{eq:drhodt-naive}
reveals that $H - i K$ can be interpreted as a non-Hermitian effective Hamiltonian 
for the subsystem. The additional {\it Lindblad term} in Eq.~\eqref{evol:Lindblad-2}
is necessary to preserve the trace of $\rho$.

\section{Effective Theory from Deeply Inelastic Reactions}
\label{sec:Effective Theory}

In this section, we review how effective  field theories for cold atoms
are constructed, with particular focus on the treatment of deeply inelastic
reactions. We first discuss the construction of the effective
Lagrangian, and how integrating deeply inelastic reaction products out
of a theory results in new local operators in the effective Lagrangian 
density. We then discuss how such a formalism 
can be used to study multi-particle
states by means of an effective density matrix obtained by tracing over
states that include deeply inelastic reaction products.

\subsection{Locality} 
\label{sec:Locality}
    
\begin{figure}[t]
\centerline{ \includegraphics*[width=10cm,clip=true]{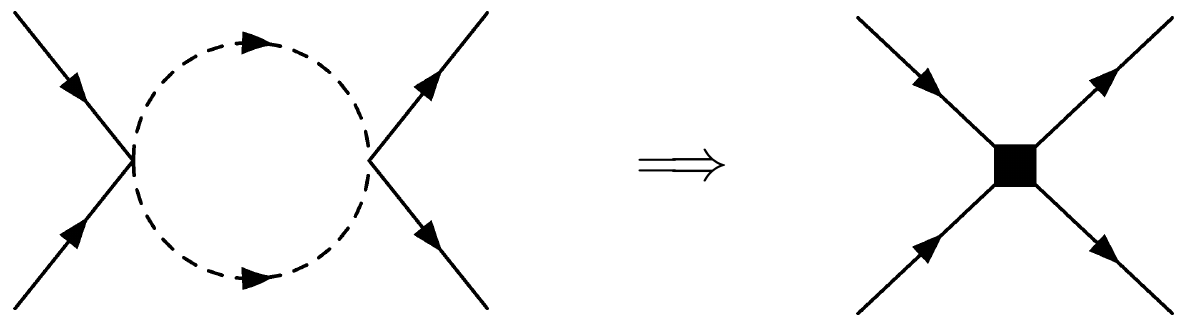} }
\caption{
The amplitude for the elastic scattering process $\psi \psi \to \psi \psi$
has contributions from intermediate states $\phi \phi$, where $\phi$ is a
different hyperfine state. If  $\phi$ is a higher hyperfine state,
the virtual atoms can propagate only over short distances,
so their effects can be mimicked by local $2 \to 2$ operators.
If  $\phi$ is a lower hyperfine state,
the virtual atoms can propagate over long distances,
but the imaginary part of the amplitude is still local
and  can again be mimicked by local $2 \to 2$ operators.
}
\label{fig:2to2-elastic}
\end{figure}

An effective field theory is obtained by removing (``integrating out'')
states from a field theory. The simplest applications involve removing
particles with much higher energies. 
We consider atoms with mass $M$ in a specific hyperfine state labelled $\psi$
and with kinetic energies much smaller than the splitting $\Delta$ between hyperfine states.
As illustrated in Fig.~\ref{fig:2to2-elastic}, 
there is a contribution to the elastic scattering amplitude from scattering 
into a virtual pair of particles in a higher hyperfine state labelled $\phi$,
which then rescatter into atoms in the original hyperfine state.
The virtual atoms are off their energy shells by large amounts of order $\Delta$.
By the time-energy uncertainty principle, the virtual states have short lifetimes of order $\hbar/\Delta$,
during which the $\phi$ atoms can propagate only over short distances 
much smaller than the wavelengths of the $\psi$ atoms.
The effects of the higher hyperfine states can be modeled in the effective field theory
by a contact interaction of the form $\psi^\dagger \psi^\dagger \psi \psi$, 
where $\psi$ is the quantum field that annihilates an atom in the specific hyperfine state $\psi$.
(We use the same symbol for the label of the hyperfine state and for its quantum field.)
The $2 \to 2$ contact interaction is the leading term in a series of
local operators in the effective Lagrangian that mimic the  effects of the higher hyperfine states
to arbitrary precision. This series is obtained at tree-level by
Taylor expanding the scattering amplitude in the momentum transfer
and then writing down terms in the effective Lagrangian density that reproduce this expansion:
\begin{equation}
  \delta\Lag = 
  g\, \psi^\dagger \psi^\dagger \psi \psi
  + h \,\nabla(\psi^\dagger\psi) 
  \cdot \nabla(\psi^\dagger\psi) + \ldots,
  \label{eq:deltaL}
\end{equation}
where  the coefficients $g$, $h$, \ldots\ are real valued.
The individual operators are renormalized
when loop corrections are included, but the effective theory
is still capable of reproducing the original theory to arbitrary
precision provided operators with a sufficient number of gradients are
retained. Operators with $n$ gradients correct the theory at
order $(q/P)^n$, where $q$~is the momentum transfer and
$P = (M \Delta)^{1/2}$ is the momentum scale associated with the 
energy splitting $\Delta$ between the hyperfine states.
We have integrated the higher hyperfine states out of the theory by replacing them 
by the interactions terms in Eq.~\eqref{eq:deltaL}.
   
\begin{figure}[t]
\centerline{ \includegraphics*[width=15cm,clip=true]{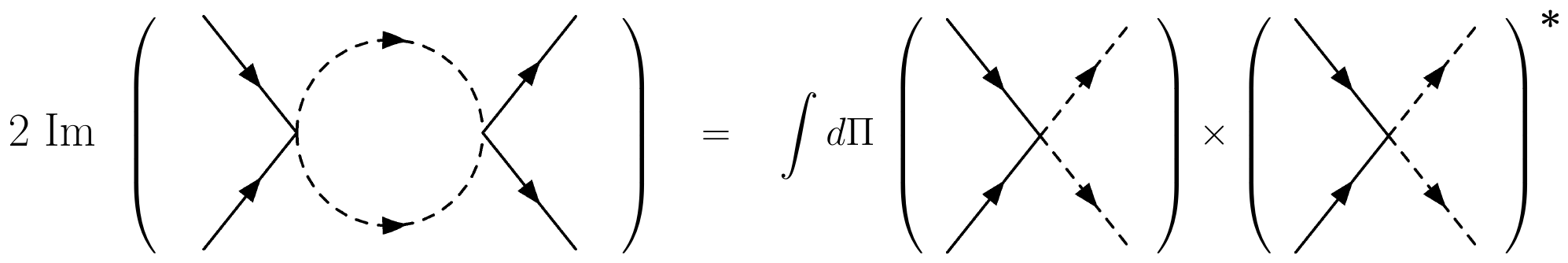} }
\caption{
The optical theorem relates the imaginary part of the 
    amplitude for $\psi \psi \to \phi \phi \to \psi \psi$,
    where $\phi$ is a lower hyperfine state,
to the rate for the deeply inelastic scattering process $\psi \psi \to \phi \phi$.
}
\label{fig:2to2-inelastic}
\end{figure}

A less obvious opportunity to remove states from the theory arises when atoms in the hyperfine state $\psi$
scatter inelastically into atoms in a  lower hyperfine state $\phi$. 
Such a scattering process is an example of a \emph{deeply inelastic reaction}, 
in which the difference in the rest energies of the initial-state and final-state atoms is
converted into large kinetic energies of the final-state atoms.
The rate for the deeply inelastic scattering process $\psi \psi \to \phi \phi$ is 
related by the optical theorem to the imaginary part of the 
amplitude for $\psi \psi \to \phi \phi \to \psi \psi$ (see
Fig.~\ref{fig:2to2-inelastic}). That amplitude is analytic in the momentum transfer $\bm{q}$,
because the nearest non-analyticity in the amplitude is at
the threshold energy for the $\phi \phi$~state,
which is lower by an amount of order $ \Delta$. As a result, we can Taylor expand the amplitude in
powers of~$\bm{q}/(M \Delta)^{1/2}$ and reproduce this expansion by terms in the
effective Lagrangian as in Eq.~\eqref{eq:deltaL}.
In this case, the coefficients $g$, $h$, \ldots\ are complex valued.
We are particularly interested in their imaginary parts,
which come from the deeply inelastic scattering reaction.
These terms mimic the effects of the lower hyperfine states in the effective theory.
       
It may seem nonintuitive
that the effects of inelastic scattering to on-shell particles
can be mimicked by local operators, because once an atom in the lower hyperfine state $\phi$ 
is created, it can propagate over long distances. In fact, as far as low-energy atoms 
in the hyperfine state $\psi$ are concerned, the inelastic scattering  process is quite
local. This is because the reaction region in which the low-energy  atoms disappear can be reconstructed
by tracking the inelastic scattering  products back to their origin. The inelastic reaction products
have high momenta of order~$(m \Delta)^{1/2}$ and therefore
short wavelengths of order~$\hbar/(m \Delta)^{1/2}$,  so they
can locate the decay with a resolution of order~$\hbar/(m \Delta)^{1/2}$. 
Therefore the reaction is localized over a
region of size~$\hbar/(m \Delta)^{1/2}$, which is very small compared to the
wavelengths of the incoming low-energy atoms.

\subsection{Effective density matrix for decaying atoms}

Some aspects of the effective field theory obtained by integrating out deeply inelastic reactions 
are most easily understood by considering a deeply inelastic 1-body reaction,
namely the decay of an atom.
The atom could be a metastable excited state, such as the $2 ^3S_1$ state of $^4$He,
which was one of the earliest atoms for which Bose-Einstein condensation was achieved  
\cite{Aspect:2001,Cohen-Tannoudji:2001}. The radiative decay of the atom into its ground state
is a deeply inelastic reaction. For simplicity, we 
assume the ground-state atoms interact weakly with the metastable atoms
and that, once they are produced, the ground-state atoms quickly escape from  the system.
The Hamiltonian $H - iK$ for the effective theory is the sum of the Hermitian
Hamiltonian for the metastable atoms 
and an anti-Hermitian piece $-iK$.
The Hermitian operator $K$ can be expressed as $\frac12 \Gamma\, N$,
where $\Gamma$ is the decay rate of the metastable atom 
and $N$ is the number operator that counts the metastable atoms: 
\begin{equation}
  N = \int d^3\rv \, \psi^\dagger \psi .
  \label{eq:Nmuhat}
\end{equation}
    
The quantum mechanics of such a theory is unconventional.
The effective Hamiltonian $H - iK$ commutes with the number operator $N$,
so the time evolution generated by $H - iK$ does \emph{not}
change the number of atoms in a state.
Instead the effects of atom decay 
must be taken into account by transferring
the probability carried by the $n-$atom component 
to states containing fewer atoms. An $n-$atom state
evolves into a superposition of states with $n$ and fewer atoms.
The norm of a state containing $n$~atoms
decreases to zero exponentially with the decay rate~$n \Gamma$. 
This is the correct result:  if the probability for one atom to
still be an atom after
time~$t$ is $\exp(-\Gamma t)$, the probability for $n$~atoms to still
be $n$~atoms is $\exp(-n\Gamma t)$.

We typically want more information about where the probability goes. An $n-$atom state
evolves into a superposition of states with $n$, $n-1$, $n-2$\,\ldots\,atoms.
In the full theory,  the superposition can be described by the density matrix $\rho_\full$. 
In the effective theory, we
describe the superposition of states with different atom numbers 
using an \emph{effective density matrix} $\rho$, from which
we remove ground-state atoms
by tracing over states that include them:
\begin{equation}
  \label{eq:rho-eff-defn}
  \rho \equiv \Tr_\deep \left(\rho_\full \right).
\end{equation} 
The subscript ``deep'' indicates that the partial trace is over states that include 
deeply inelastic reaction products.
This effective density matrix, like $\rho_\full$, is
Hermitian, positive, and has unit trace: $\Tr(\rho)=1$. These
properties follow from the definition in Eq.~(\ref{eq:rho-eff-defn}).
Naively we might expect  the time evolution equation for $\rho$ to be
Eq.~\eqref{eq:drhodt-naive}, which reduces to
\begin{equation}
  i\frac{d\ }{dt}\rho =   \left[H, \rho\right]
  - \frac{i}{2} \Gamma \left\{ N, \rho\right\} .
\end{equation}
However this equation does \emph{not} conserve
the total probability~$\Tr(\rho)$.
The correct evolution equation is the Lindblad equation:
\begin{equation}
  i\frac{d}{dt}\rho =
  \left[H, \rho\right]
  - \frac{i}{2} \Gamma \left\{ N, \rho\right\}
  + i \Gamma \int d^3\rv\,\psi(\rv)\,\rho\,\psi^\dagger(\rv).
  \label{eq:muon-evol-eq}
\end{equation}
Time evolution preserves the trace of $\rho$, because the trace of
the additional Lindblad term cancels the trace of the anticommutator 
term in Eq.~\eqref{eq:muon-evol-eq}.
    
The role of the Lindblad term is easily understood
if we use the evolution equation
to calculate the rate of change of the
probability $P_n(t)$ for finding $n$~metastable atoms in the system. 
This probability can be expressed as the partial trace of $\rho$
over states $ | X_n \rangle$ that contain $n$ atoms:
\begin{equation}
  P_n(t) \equiv \sum_{X_n} \langle X_n | \rho(t) | X_n \rangle.
\end{equation}
The partial trace of the  evolution equation in Eq.~\eqref{eq:muon-evol-eq} gives
\begin{equation}
  \frac{d\ }{dt}  P_n(t)= - n\Gamma P_n(t) + (n+1)\Gamma P_{n+1}(t).
  \label{eq:dPn(t)}
\end{equation}
The partial trace of the commutator term in Eq.~\eqref{eq:muon-evol-eq} is 0,
because the operator $H$ does not change the atom number.
This allows a complete set of $n-$atom states to be inserted between $H$ and $\rho$. 
The partial trace of the anticommutator term in Eq.~\eqref{eq:muon-evol-eq}
gives $-n\Gamma P_n$, which is the rate at which
probability leaves the $n-$atom sector because of 
the decay of an atom. The partial trace of the Lindblad
term in Eq.~\eqref{eq:muon-evol-eq} gives $+(n+1)\Gamma P_{n+1}$,
which is the rate at which probability enters the $n-$atom sector
from the decay of an atom in the $(n+1)-$atom sector. 
This expression can be obtained by inserting complete sets of $(n+1)-$atom states
on the left and right of $\rho$ in the Lindblad term in Eq.~(\ref{eq:muon-evol-eq})
and then rearranging the factors to give
\begin{eqnarray}
  i \Gamma \sum_{X_{n+1}} \sum_{X'_{n+1}}\,
  \langle X'_{n+1}| \rho |X_{n+1} \rangle 
  \langle X_{n+1} | N |X'_{n+1} \rangle.
  \label{eq:drho-lindblad}
\end{eqnarray}
The number operator $N$, which is given in Eq.~\eqref{eq:Nmuhat},
was obtained by replacing the sum of $ |X_n \rangle \langle X_n |$
between $ \psi^\dagger(\rv)$ and $\psi(\rv)$ by the identity operator.
The number operator in Eq.~\eqref{eq:drho-lindblad} can be replaced by its eigenvalue $n+1$.

The Lindblad term in Eq.~(\ref{eq:muon-evol-eq}) is essential to get the correct physical behavior 
for the time evolution of the total number of metastable atoms. The expectation value of the atom number is
\begin{equation}
  N(t) \equiv {\rm Tr} \big( N \rho(t)\big) = \sum_n n P_n(t).
  \label{eq:Nmu-t}
\end{equation}
We can use Eq.~\eqref{eq:dPn(t)} to determine the time dependence of $N(t)$: 
\begin{equation}
  \frac{d\ }{dt} N(t)= -\Gamma \Big[
    \sum_n n^2 P_n(t) - \sum_n n(n+1) P_{n+1}(t) \Big] .
\end{equation}
After shifting the index of the second term on the right side,
we obtain
\begin{equation}
  \frac{d\ }{dt} N(t)= -\Gamma N .
\end{equation}
This implies that $N(t) = N_0 \exp(-\Gamma t)$, as expected for  a collection of 
atoms with decay rate $\Gamma$.

\subsection{Effective density matrix from deeply inelastic reactions}

We now consider a more general effective field theory obtained by integrating out the
reaction products from deeply inelastic reactions.
By the arguments presented in Sec.~\ref{sec:Locality},
the effects of the deeply inelastic reactions
can be reproduced by local anti-Hermitian terms in the effective Hamiltonian. 
The effective Hamiltonian can be expressed in the form $H_\eff = H  - i K$,
where $H$ and $K$ are both  Hermitian operators.
The operator  $K$ can be expressed in  the form
\begin{equation}
K = \sum_i \gamma_i \int d^3r\,  \Phi_i^\dagger \Phi_i ,
\label{eq:K-Phi}
\end{equation}
where the local operators $\Phi_i(\bm{r})$  annihilate low-energy atoms
in configurations  corresponding to inelastic reactions.
The real constants $\gamma_i$ can be determined by
matching low-energy observables in the full theory and in the effective theory.

The Hamiltonian $H  - i K$ commutes with the number operator $N$ for low-energy atoms,
so the time evolution generated by $H - iK$ does \emph{not} change the number of low-energy atoms in a state.
Instead the effects of the deeply inelastic reactions 
must be taken into account by transferring
the probability carried by the $n-$atom component 
to states containing fewer atoms.
The norm of a state containing $n$~atoms
decreases  to zero at a rate that increases with $n$.
An $n-$atom state
evolves into a superposition of states with $n$ and fewer atoms.
We describe this superposition of states using an \emph{effective density matrix} $\rho$
defined by tracing over the deeply inelastic decay products
as in Eq.~\eqref{eq:rho-eff-defn}.   More precisely,
we trace out any state containing an atom with momentum
exceeding the ultraviolet cutoff of the effective field theory.
The effective density matrix, like the underlying density matrix, is
Hermitian, positive, and it has unit trace: $\Tr(\rho)=1$. These
properties follow from the definition in Eq.~(\ref{eq:rho-eff-defn}).

As is familiar in the field of quantum information \cite{Preskill}, the definition
of the effective density matrix $\rho$  as the partial trace in Eq.~(\ref{eq:rho-eff-defn})
implies that $\rho(t)$ satisfies a self-contained time evolution equation 
that is completely determined by $\rho$ 
and by the full density matrix $\rho_\full(0)$ at some initial time.
This evolution equation is however non-Markovian:
the time derivative $(d/dt)\rho(t)$ is determined not only
by $\rho(t)$ but also by $\rho(t')$ at the earlier times $0<t'<t$.
This dependence on the history takes into account the effects
of high-momentum atoms that are produced by inelastic reactions 
and subsequently interact with the low-energy atoms.
However the density matrix $\rho(t)$ for the effective theory
does not need to reproduce the full density matrix 
on short time scales of order  $\hbar /E_{\rm deep}$.
It is sufficient to reproduce its effects after  time averaging over time intervals 
much larger than $\hbar /E_{\rm deep}$.
This time average removes transients associated with high-momentum atoms 
that cannot be described accurately within the effective field theory.
It is only after removing these transients that 
it is possible to have a density matrix $\rho(t)$ that is Markovian. 
Thus the proper definition of the effective density matrix 
requires that
the partial trace in Eq.~(\ref{eq:rho-eff-defn})
be supplemented by an appropriate time average
that makes $\rho(t)$ Markovian. 

Naively we might expect  the time evolution equation for $\rho$ to be Eq.~\eqref{eq:drhodt-naive},
where $K$ is the Hermitian operator in Eq.~\eqref{eq:K-Phi}.   
However, this equation does \emph{not} conserve  the total probability~$\Tr(\rho)$.
The correct evolution equation is the Lindblad equation:
\begin{equation}
i \hbar \frac{d\ }{dt} \rho = \left[H, \rho\right]- i \left\{K,\rho\right\}
+2i \sum_i \gamma_i   \int \!\!d^3r\, 
\Phi_i \rho \Phi_i^\dagger.
\label{eq:LindbladPhi}
\end{equation}
As in Eq.~\eqref{eq:muon-evol-eq},
the trace of the additional  Lindblad term cancels the trace of the anticommutator term.
The local operators $\Phi_i(\bm{r})$, which  annihilate
configurations of low-energy atoms that can undergo a deeply inelastic reaction,
are the {\it Lindblad operators}. In quantum information theory,
a Lindblad operator is sometimes called a {\it quantum jump operator}.
In the low-energy effective theory, a Lindblad operator produces a jump to
a state with a different low-energy atom number.
Given the form of~$K$ in Eq.~\eqref{eq:K-Phi}, the Lindblad equation is a necessary
consequence of our physical requirements
on the effective density matrix: $\rho$~is Hermitian, positive,
has unit trace, and it is Markovian. 
An explicit diagrammatic illustration of the emergence of the
Lindblad equation from tracing over 
states that include deeply inelastic reaction products
is given in Appendix~\ref{sec:Deductive}.
    
In order to obtain the Lindblad equation in Eq.~\eqref{eq:LindbladPhi},
it is essential that $K$ have the structure shown
in~Eq.~(\ref{eq:K-Phi}). A more generic form for this operator is
\begin{equation}
  K = \int \!\!d^3 r \sum_{ij} c_{ij} \Phi_i^\dagger(\rv) \Phi_j(\rv),
  \label{eq:K-structure-gen}
\end{equation}
where $c_{ij}$ is a positive Hermitian matrix. 
It is a Hermitian matrix, because $K$ is Hermitian by definition. 
It is guaranteed to be a positive matrix by the optical theorem,
which implies that the $T$ matrix satisfies $-i (T - T^\dagger) = T^\dagger T$.
Matrix elements of $K$ reproduce
the anti-Hermitian parts of scattering amplitudes $\bra{b}T\ket{a}$,
where~$\ket{a}$ and~$\ket{b}$ are states in the effective theory
that are connected by intermediate deeply inelastic  reaction channels.
The Hermitian parts of these amplitudes are reproduced by $H$.
The optical theorem guarantees the positivity of the anti-Hermitian parts.
The double sum in Eq.~\eqref{eq:K-structure-gen} is easily rewritten in the canonical
form in Eq.~\eqref{eq:K-Phi} by expanding $c_{ij}$ in terms of outer products of its eigenvectors.
 
\section{Atom Loss from Deeply Inelastic Reactions}
\label{sec:AtomLoss}

In this section, we use the Lindblad equation to determine the time evolution 
of the mean number of low-energy atoms, which
can be changed by deeply inelastic reactions.
We also derive universal relations for the inelastic two-atom loss rate 
for fermions with two spin states
and for the inelastic three-atom loss rate for identical bosons.

\subsection{Mean Atom Number}

We consider an open effective field theory with effective Hamiltonian
$H_{\rm eff}=H-iK$, 
where $H$ and $K$ are Hermitian.
The Hermitian operator $K$ in the anti-Hermitian part can be expressed 
in terms of local Lindblad operators $\Phi_i(\bm{r})$ as in
Eq.~\eqref{eq:K-Phi}.
The time evolution of the density matrix $\rho$ for the effective field theory
is given by the Lindblad equation in Eq.~\eqref{eq:LindbladPhi}.

We reconsider  an atom number operator $N$ 
that is conserved by the reactions  in the effective field theory
but is violated by deeply inelastic reactions in the full theory.
Since Lindbladian time evolution preserves the trace of $\rho$, 
the system average of $N$ can be expressed as
\begin{equation}
\langle N \rangle = {\rm Tr}(\rho N).
\label{N-ave}
\end{equation}
The statement that $N$ is conserved by the reactions in the effective theory
can be expressed as the commutation relation
\begin{equation}
\big[ N,H\big ]=0.
\label{[N,H]}
\end{equation}
The inelastic reaction responsible for the $\Phi_i^\dagger \Phi_i$ term
in $K$ changes the atom number $N$ by some integer $k_i$.  
This statement can be expressed as a commutation relation:
\begin{equation}
\big[ N,\Phi_i(\bm{r}) \big]=-k_i \Phi_i(\bm{r}).
\label{[N,Phi]}
\end{equation}

We now consider the time evolution of $\langle N \rangle$.
By taking the time derivative of ${\rm Tr}(\rho N)$,
inserting the time evolution equation in Eq.~\eqref{eq:LindbladPhi},
and rearranging the terms,
we get
\begin{eqnarray}
\frac{d\ }{dt} \langle N \rangle =
- \frac{i}{\hbar} {\rm Tr}(\rho [N,H])
 - \frac{1}{\hbar} \sum_i \gamma_i  \int \!\! d^3r \,
{\rm Tr} \big( \rho (N\Phi_i^\dagger \Phi_i  +\Phi_i^\dagger \Phi_i  N
-2 \Phi_i^\dagger N \Phi_i) \big).
\label{Nevol:Lindblad}
\end{eqnarray}
The first term on the right side vanishes, 
because $N$ commutes with $H$.
By using the  commutation relation in Eq.~\eqref{[N,Phi]}
in the second term on the right side, the rate of change of atom number
reduces to
\begin{eqnarray}
\frac{d\ }{dt} \langle N \rangle &=& -\frac{2}{\hbar} 
\sum_i k_i \gamma_i \int \!\! d^3r \, \langle \Phi_i^\dagger \Phi_i \rangle.
\label{Nevol:Phi}
\end{eqnarray}
The rate of change in $\langle N \rangle$ has been expressed as
a sum of expectation values of the same terms that appear in 
the expression for $K$ in Eq.~\eqref{eq:K-Phi} but multiplied by integers $ k_i$. 

\subsection{Three-Body Recombination Rate}
\label{sec:3bodyrecombination}

The atoms that are studied in cold atom experiments
can form deeply bound diatomic molecules whose binding energies are 
orders of magnitude larger than the energy scales of the cold trapped atoms.
Three-body recombination, in which three low-energy atoms collide 
and two of them bind into a molecule, is
a deeply inelastic reaction if the molecule is deeply bound.
The momenta of the outgoing molecule and the recoiling atom are often
large enough for them to escape from the trapping potential.
In a locally homogeneous gas of atoms in thermal equilibrium,
the rate of decrease in the local number density $n_A$ of low-energy atoms 
can be expressed as a rate equation:  
\begin{equation}
(d/dt) n_A = - 3 K_3(T) n_A^3,
\label{dn/dt-thermal}
\end{equation}
where $K_3(T)$ is the 3-body recombination event rate constant,
which depends on the temperature $T$.

Bosonic atoms that are all in the same hyperfine state
can be described by an effective field theory with a quantum field $\psi$.
At the  low temperatures in cold atom experiments, 
the only relevant interaction between the atoms is 
S-wave scattering, whose strength is determined by the scattering length $a$.
The interaction term in the Hamiltonian  is
\begin{equation}
H_{\rm int} = (2 \pi \hbar^2 a/m) \int d^3 r\, \psi^{\dagger 2}\psi^2.
\label{Hint-psi}
\end{equation}
This interaction term can be treated using perturbation theory,
provided the scattering length $a$ is not much larger than the range
of the interactions between the atoms, 
which for ultracold atoms is the van der Waals length scale.

We now consider a deeply inelastic three-body recombination reaction 
that  produces a diatomic molecule with binding energy much larger than the temperature $T$.
Its effect on low-energy atoms  can be described in the effective field theory
by adding to the effective Hamiltonian the anti-Hermitian term $-i K$,
where the local Hermitian operator $K$ is
\begin{equation}
K = (\hbar K_3/12) \int d^3 r\,(\psi^3)^\dagger  \psi^3,
\label{Kint-psi3}
\end{equation}
and where $K_3$ is a constant that does not depend on the temperature.
The expression for $K$ in Eq.~\eqref{Kint-psi3}
has a single term with Lindblad operator $\Phi_i = \psi^3$
and coefficient $\gamma_i = \hbar K_3/12$.
The expression for the atom loss rate in Eq.~\eqref{Nevol:Phi}
thus has a single term with integer $k_i=3$. In a locally
homogeneous system of atoms, the local version of the loss rate in Eq.~\eqref{Nevol:Phi} is
\begin{eqnarray}
\frac{d\ }{dt} \langle \psi^\dagger  \psi \rangle &=& 
- \frac{K_3}{2} \,\langle (\psi^3)^\dagger  \psi^3 \rangle.
\label{Nevol:Phi3}
\end{eqnarray}
The expectation value of $\psi^\dagger  \psi $ is the local number density:
$n_A = \langle \psi^\dagger  \psi \rangle$. In a thermal gas,  
the expectation value of $(\psi^3)^\dagger  \psi^3$ is $6 n_A^3$.
Comparing with Eq.~\eqref{dn/dt-thermal},
we see that the constant $K_3$ in Eq.~\eqref{Kint-psi3} is the 
$T \to 0$ limit of the 3-body recombination event rate constant $K_3(T)$.
Its temperature dependence is negligible, 
because $kT$ is much smaller than the binding energy of the molecule.
In a Bose-Einstein condensate at zero temperature, 
the expectation value of $(\psi^3)^\dagger  \psi^3$ can be expressed as 
$|\langle  \psi \rangle^3|^2 =  n_A^3$,
where  $\langle  \psi \rangle$ is the mean field.
The atom loss rate is given by Eq.~\eqref{dn/dt-thermal}
except that the prefactor on the right side is smaller than for a thermal gas by
a factor of 1/6.

\subsection{Inelastic Two-Atom Loss Rate}
\label{sec:2atomloss}

In cold atom experiments,
the scattering length $a$ can be controlled experimentally by tuning an external  
magnetic field to near a Feshbach resonance.
If $a$ is much larger than the range of the interactions between the atoms, 
which for ultracold atoms is the van der Waals length scale,
the interactions must be treated  nonperturbatively.

Fermionic atoms in two hyperfine states 1 and 2 with scattering length $a$
can be described by  an effective field theory 
with quantum fields $\psi_1$ and $\psi_2$.
The interaction term in the Hamiltonian can be expressed as
\begin{equation}
 H_{\rm int} = \frac{g_0}{m}\int d^3 r\,
\big(\psi_2 \psi_1\big)^\dagger \big( \psi_2 \psi_1 \big),
\label{Hint-psi0}
\end{equation}
where $g_0$ is the bare coupling constant.
If the ultraviolet cutoff $\Lambda$ is imposed on the
momenta of virtual atoms, the bare coupling constant is
\begin{equation}
g_0 = \frac{4 \pi}{1/a-(2/\pi)\Lambda}.
\label{g0}
\end{equation}

If the pair of atoms in the hyperfine states 1 and 2 has a {\it spin-relaxation} scattering channel 
into a pair atoms in lower hyperfine states 3 and 4,
the optical theorem implies 
that the scattering length $a$ is complex with a small negative imaginary part.
The energy release from the spin-relaxation reaction
is much larger than the energy scales for ultracold atoms,
so this is a deeply inelastic reaction. 
The high-momentum atoms in the hyperfine states 3 and 4 can be integrated out to get 
an effective field theory for low-energy  atoms in the hyperfine states 1 and 2 only.
The interaction term in the Hamiltonian is still that in Eq.~\eqref{Hint-psi},
except that now $H_{\rm int}$ has an anti-Hermitian part
because the complex scattering length $a$ 
makes the bare coupling constant $g_0$ in Eq.~\eqref{g0} complex.

Determining the loss rate of atoms due to the deeply inelastic spin-relaxation
reaction is not trivial, because the large scattering length makes the problem nonperturbative.
However, an exact result for the inelastic two-atom loss rate
in any state has been proposed by Shina Tan \cite{Tan-unpub}.
The rates of decrease in the numbers $N_1$ and $N_2$ of atoms in the two hyperfine states are
given by the universal relation
\begin{equation}
\frac{d\ }{dt} \langle N_1 \rangle= \frac{d\ }{dt} \langle N_2 \rangle =
- \frac{\hbar}{2 \pi m}{\rm Im} \big( 1/a \big)\, C,
\label{uni2loss}
\end{equation}
where $C$ is  a property of the system called the {\it contact}.
The coefficient of $C$, which is proportional to the imaginary part of $1/a$,
is determined by 2-body physics.
The contact was first introduced by Shina Tan in 2005 \cite{Tan:0505,Tan:0508}.
It is the  thermodynamic variable conjugate to $1/a$.            
The contact $C$ is an extensive variable:
it is the integral over space of the contact density  ${\cal C}$,         
which measures the number of 1-2 pairs per (volume)$^{4/3}$.  
(The unusual power of volume arises from an anomalous dimension
associated with a non-Gaussian renormalization group fixed point \cite{Braaten:2010if}.)
Shina Tan derived other  universal relations involving the contact
that hold for any state of the system  \cite{Tan:0505,Tan:0508}.
They include the ``adiabatic relation'' 
that specifies how the free energy of a system
is affected by a change in the scattering length \cite{Tan:0508}:
\begin{equation}
\frac{d~~~~}{d(1/a)}F =
- \frac{\hbar^2}{4 \pi m} C.
\label{adiabatic}
\end{equation}
Many other universal relations involving the contact were subsequently
derived \cite{Braaten:2010if}.

The universal relation  for the inelastic two-atom loss rate in Eq.~(\ref{uni2loss}) 
can be derived by expressing the Hermitian operator $K$
in the effective Hamiltonian $H-iK$ as
\begin{equation}
K = \frac{\hbar^2}{4\pi m}{\rm Im} \big( 1/a \big) 
\int\! d^3r\, \Phi^\dagger \Phi,
\label{Kint-psi}
\end{equation}
where the local Lindblad operator is
\begin{equation}
\Phi = g_0 \psi_2 \psi_1.
\label{Phi-psi2}
\end{equation}
The operator product $ \psi_2 \psi_1$ is singular:  
its matrix elements diverge linearly with the cutoff $\Lambda$.
The multiplication by $g_0$, 
which according to Eq.~\eqref{g0} decreases as $1/\Lambda$ for large $\Lambda$,
makes $\Phi$ a finite operator
whose matrix elements have well-behaved limits as $\Lambda \to \infty$.
In a spin-relaxation scattering reaction, the decreases in the atom numbers 
$N_1$ and $N_2$ are both $k_i=1$.  According to Eq.~\eqref{Nevol:Phi},
the rates of decrease in their system averages  are therefore 
\begin{eqnarray}
\frac{d\ }{dt} \langle N_1 \rangle= \frac{d\ }{dt} \langle N_2 \rangle =
- \frac{\hbar}{2 \pi m}{\rm Im} \big( 1/a \big) 
\int\! d^3r\, \left\langle \Phi^\dagger \Phi \right\rangle.
\label{N12evol:psi}
\end{eqnarray}
The universal loss rate in Eq.~(\ref{uni2loss}) then follows from the identification
of the contact density as  \cite{Braaten:2008uh}
\begin{eqnarray}
{\cal C} =
\left\langle\Phi^\dagger \Phi \right\rangle.
\label{contactdensity}
\end{eqnarray}
This identification can be verified by using the adiabatic relation in Eq.~\eqref{adiabatic}.
The only dependence of the Hamiltonian on the scattering length $a$
is in the  interaction term in the Hamiltonian density ${\cal H}_{\rm int}$ in Eq.~\eqref{Hint-psi}.
If the tiny imaginary part of the scattering length is neglected,
the derivative of the Hermitian part $H$ of the effective Hamiltonian with respect to $1/a$ is
\begin{equation}
\frac{d~~~~}{d(1/a)}H = - \frac{\hbar^2}{4 \pi m} \, 
\int\! d^3r\, \Phi^\dagger \Phi.
\label{dHint/da}
\end{equation}
By the Feynman-Hellman theorem, the system average of the left side of this equation 
is the left side of the adiabatic relation in Eq.~\eqref{adiabatic}.
With the identification of the contact density in Eq.~\eqref{contactdensity},
the system average of the right side of Eq.~\eqref{dHint/da} 
is the right side of the adiabatic relation in Eq.~\eqref{adiabatic}.
This completes the derivation of the inelastic two-atom loss rate  in Eq.~(\ref{uni2loss}) 
and the adiabatic relation in Eq.~\eqref{adiabatic}.

In Ref.~\cite{Braaten:2013eya},
the authors presented an incorrect derivation of the universal 
relation for the inelastic two-atom loss rate
that suggests that there are additional terms on the right side of Eq.~\eqref{uni2loss}.
They assumed  the density matrix $\rho$ satisfies the naive time evolution equation in
Eq.~\eqref{eq:drhodt-naive}.  This equation implies that the number of atoms is conserved,
but that the probability ${\rm Tr}(\rho)$ decreases with time.
In Ref.~\cite{Braaten:2013eya}, the authors assumed that  
the mean atom number $\langle N \rangle$ 
was given by ${\rm Tr}(\rho N)/{\rm Tr}(\rho)$.
The resulting expression for the time derivative of $\langle N \rangle$
can be expressed as a double integral over space of an expression
involving the product of the number density operator and the Hamiltonian density operator.
Upon applying the operator product expansion to the product of these operators,
they obtained an expansion  for $(d/dt)\langle N \rangle$
in terms of increasingly higher-dimension operators
with the leading term given by Eq.~\eqref{uni2loss}.
The derivation of the universal relation  using the
Lindblad equation makes it clear that there are no additional terms
beyond the contact term in Eq.~(\ref{uni2loss}).

\subsection{Inelastic Three-Atom Loss Rate}
\label{sec:3atomloss}

Bosonic atoms that are all in the same hyperfine state
can be described by a nonrelativistic quantum field theory with a quantum field $\psi$ \cite{Braaten:2004rn}.
If the scattering length $a$ is much larger than the range of interactions of the atoms,
two-atom interactions and three-atom interactions must both be  treated nonperturbatively.
The resulting quantum field theory is characterized by a discrete scaling symmetry 
in which lengths and time are multiplied by integer multiples of $\lambda_0$ and $\lambda_0^2$,
respectively, where  $\lambda_0= e^{\pi/s_0}$ and $s_0\approx1.00624$ is a universal number.
Three-atom interactions are determined by a parameter $\Lambda_*$ with dimensions of momentum,
upon which physical observables can only depend log-periodically.
If $a$ is positive, two-atom interactions produce a weakly bound diatomic molecule  called the {\it shallow dimer}.  
Regardless of the sign of $a$, three-atom interactions produce a sequence of 
weakly bound triatomic molecules called {\it Efimov trimers} \cite{Braaten:2004rn}.
In Ref.~\cite{Braaten:2011sz}, several universal relations were derived that hold for any system consisting
of identical bosons.
The universal relations relate various properties of the system to two extensive thermodynamic variables:
the 2-body contact $C_2$, which is the analog of the contact for  fermions with two spin states,
and the 3-body contact $C_3$.

The atoms that are studied in cold atom experiments
form deeply bound diatomic molecules ({\it deep dimers}) whose binding energies are 
orders of magnitude larger than the energy scales of the cold trapped atoms.
The deep dimers provide various pathways for deeply inelastic reactions
that result in the loss of three atoms.  They include (a)
three-body recombination, in which three low-energy atoms collide 
and two of them bind into a deep dimer,
(b) dimer relaxation, in which a shallow dimer and an atom scatter into a deep dimer and an atom,
and (c) Efimov trimer decay into a deep dimer and an atom.
In Ref.~\cite{Smith:2013eoa}, a universal relation for the three-atom inelastic loss rate
was presented but not derived.
In this section, we use the Lindblad equation to sketch the derivation of this relation. 

To derive the universal relation for the three-atom inelastic loss rate,
we follow the formalism laid out in Ref.~\cite{Braaten:2011sz}.
In the absence of deep dimers, the interaction term in the Hamiltonian can be written as
\begin{equation}
H_{\rm int} = \frac{g_2}{4m} \, \int\! d^3r\, (\psi^2)^\dagger\psi^2
+\frac{g_3}{36m} \, \int\! d^3r\, (\psi^3)^\dagger\psi^3.
\label{Hint-psi-3b}
\end{equation}
Since this effective field theory has ultraviolet divergences, it
must be regularized.  The bare coupling constants $g_2$ and $g_3$ depend on
an ultraviolet momentum cutoff $\Lambda$:
\begin{subequations}
\begin{eqnarray}
g_2(\Lambda) &=& \frac{8 \pi}{1/a-(2/\pi)\Lambda}\,,
\label{eq:g2}
\\
g_3(\Lambda) &=& h_0 \,\frac{9g_2^2}{\Lambda^2}\,
\frac{\sin(s_0 \ln(\Lambda/\Lambda_*)-\arctan(1/s_0))}
{\sin(s_0 \ln(\Lambda/\Lambda_*)+\arctan(1/s_0))} ,
\label{eq:g3}
\end{eqnarray}
\end{subequations}
where $a$ is the two-body scattering length of the bosons
and $\Lambda_*$ is a three-body parameter
introduced in Ref.~\cite{Bedaque:1998kg}. 
The three-body parameter $\Lambda_*$ can be determined
from any three-body datum, such as the binding energy 
of an Efimov trimer or the atom-dimer scattering length. 
(See Ref.~\cite{Braaten:2004rn} for explicit relations).
The numerical prefactor $h_0$ in Eq.~\eqref{eq:g3} depends on the ultraviolet
cutoff prescription and has the value $h_0\approx 0.879$
if $\Lambda$ is a sharp momentum cutoff \cite{Braaten:2011sz}.

Three-atom losses from deeply inelastic reactions 
involving deep dimers can be taken into account by adding an imaginary part to the 
coupling constant $g_3$ in the Hamiltonian in Eq.~\eqref{Hint-psi-3b}.  
The resulting anti-Hermitian part of the Hamiltonian is $-i K$, where $K$ is
\begin{equation}
K = -\frac{{\rm Im}(g_3)}{36m}
\int\! d^3r\, (\psi^{3})^\dagger\psi^3.
\label{Hint-imag-3b}
\end{equation}
Physical observables can be expressed particularly conveniently in terms of 
$\Lambda_*$ and an additional real three-body
loss parameter $\eta_*$ \cite{Braaten:2003yc,Braaten:2004rn}
defined by analytically continuing the three-body coupling constant $g_3$ in Eq.~\eqref{eq:g3}
to complex values using the substitution
\begin{equation}
\ln\Lambda_* \to \ln\Lambda_* +i\eta_*/s_0.
\end{equation}

Inserting $k_i=3$ into the expression in Eq.~\eqref{Nevol:Phi} for the three-atom
inelastic loss rate,
we obtain the universal relation
\begin{equation}
\frac{d\ }{dt} \langle N \rangle=
- \frac{6\hbar}{m s_0}\sinh(2\eta_*) C_3,
\label{uni3loss}
\end{equation}
where $C_3$ is the three-body contact,
which can be expressed as
\begin{equation}
C_3 = f(\Lambda)  \int d^3r\, \langle (\psi^{3})^\dagger\psi^3 \rangle.
\label{eq:3bcontact}
\end{equation}  
The prefactor $f(\Lambda)$ depends on the ultraviolet cutoff \cite{Braaten:2011sz}. 
Since its precise form is not very insightful, it will not be given here. 
The three-body contact can also be defined 
in terms of the expectation value of 
the logarithmic derivative with respect to $\Lambda_*$ of the Hermitian part of the 
effective Hamiltonian in Eq.~\eqref{Hint-psi-3b} at fixed $a$~\cite{Braaten:2011sz}:
\begin{equation}
\Lambda_*\frac{\partial ~~}{\partial\Lambda_*}\langle H \rangle\bigg|_a =
-\frac{2\hbar^2}{m}C_3.
\end{equation}
The universal relation in Eq.~\eqref{uni3loss},
with the factor  $\sinh(2\eta_*)$ approximated by $2\eta_*$,
was previously given in Ref.~\cite{Smith:2013eoa}.
It was also given previously  in 
Ref.~\cite{Werner:2012aa} for the special case of an Efimov trimer state.

\section{Summary}
\label{sec:Summary}

In this paper, we have shown that integrating out the high-momentum atoms produced by
deeply inelastic reactions produces an open effective field theory
in which the time evolution of the  effective density matrix is given by the Lindblad equation.
The effective density matrix can be obtained from the density matrix of the full theory
by taking a partial trace over states that include high-momentum atoms 
and then carrying out an appropriate time average to eliminate high-frequency transients.
The Lindblad operators are local, and they can be deduced from the anti-Hermitian terms 
in the effective Hamiltonian.
The Lindblad terms in the evolution equation are essential to get the 
correct evolution equation for the mean atom number.
We used the Lindblad equation to present the first correct  derivation of the 
universal relation for the two-atom inelastic loss rate for fermionic atoms 
with two spin states that  interact through a large scattering length.
We also used the Lindblad equation to present the first derivation of the 
universal relation for the three-atom inelastic loss rate for identical bosonic atoms 
with  a large scattering length.

The Lindblad equation has many other applications to atom loss processes 
in cold atom physics.  An obvious extension is to heteronuclear
systems, for which there are two types of Efimov trimers.
Due to the smaller discrete scaling factor associated with one type of Efimov trimer,
atom loss processes in such systems have  recently been of
great theoretical and experimental interest 
\cite{Helfrich:2010yr,Tung:2014,Pires:2014,Petrov:2015}.
The Lindblad equation could be applied to fermionic atoms with two spin states 
on the upper branch of the two-atom spectrum at scattering lengths for which
three-body recombination into the shallow dimer is a
deeply inelastic reaction \cite{Shenoy:2011}.
It could also be applied to losses of dipolar atoms from three-body
recombination  into deep dimers \cite{Ticknor:2010,Wang:2011a,Wang:2011b}.

\begin{acknowledgments}
E.B.\ and H.-W.H.\ acknowledge useful discussions with Shina Tan.
The research of E.B.\ was supported in part by the Department of Energy
under grant DE-FG02-05ER15715, by the
National Science Foundation under grant PHY-1310862,
and by the Simons Foundation.
The research of H.-W.H.\ was supported in part 
by the BMBF under contract 05P15RDFN1, 
and by the Deutsche Forschungsgemeinschaft (SFB 1245).
The research of G.P.L.\ was supported in part by the National Science Foundation.
\end{acknowledgments}

\appendix

\section{Diagrammatic Illustration}
\label{sec:Deductive}

In this section,  we demonstrate how the Lindblad structure
of the evolution equation for the density matrix emerges from 
a diagrammatic analysis of a simple field theory model with a deeply
inelastic 2-body reaction. 
To reduce visual clutter,
we introduce compact notation for the integral over space
and for the integral over a momentum:
\begin{equation}
    \int_{\bm{r}} \equiv \int\! \!d^3r \,, \quad\quad
    \int_{\bm{p}} \equiv \int\! \!\frac{d^3p}{(2\pi)^3}\,.
\end{equation}
We also set $\hbar = 1$.

\subsection{Field Theory Model}

We consider a quantum field theory with two fields $\psi$ and $\phi$.
We refer to the particles annihilated by these field operators as $\psi$
atoms and $\phi$ atoms,
respectively.  The field operators satisfy canonical commutation relations if
the atoms are bosons 
and canonical anticommutation relations if they are fermions.
The  Hamiltonian for the full theory has the form
\begin{equation}
\label{eq:hamiltonian}
H_{\rm{full}} = {H}_0^\psi + {H}_0^\phi + H_\mathrm{int}\,.
\end{equation}
The free-field terms in the Hamiltonian are
\begin{subequations}
\begin{eqnarray}
H_0^\psi &=&  
 \int_{\bm{r}}  \psid(\bm{r}) \left(-\frac{\nabla^2}{2M} \right) \psi(\bm{r})\,,
\\
H_0^\phi &=& 
\int_{\bm{r}}  \phid(\bm{r}) \left(-\Delta - \frac{\nabla^2}{2M} \right)  \phi(\bm{r})\,,
\end{eqnarray}
\end{subequations} 
where $M$ is the mass of the atoms.
We have chosen the rest energies of a $\psi$ atom and a $\phi$ atom to be zero and $-\Delta$, respectively.
The on-shell energies of atoms with momentum $\bm{p}$ are
\begin{subequations}
\begin{eqnarray}
E_{\bm{p}} &=& \bm{p}^2/(2M),
 \\
\omega_{\bm{p}} &=& -\Delta + \bm{p}^2/(2M).
\end{eqnarray}
\end{subequations} 
The interaction Hamiltonian includes a term that
allows a pair of $\psi$~atoms to scatter into a pair of $\phi$~atoms:
\begin{equation}
\label{eq:Hint}
H_\mathrm{int} =
    \tfrac{1}{4}g \int_{\bm{r}}
    \left(\psi^{\dagger 2}(\bm{r})\phi^2(\bm{r})
    +\phi^{\dagger 2}(\bm{r})  \psi^2(\bm{r})\right)+\ldots\,,
\end{equation}
where the ellipses denote further local interaction terms, 
such as $\psi^{\dagger 2}\psi^2$ and $\phi^{\dagger2}\phi^2$. Their
precise form will not be needed in the following discussion.

The leading contribution to the transition amplitude for the process $\psi\psi\to\psi\psi$
from inelastic reactions with a $\phi\phi$ intermediate state
comes from the imaginary part of the one-loop diagram   in Fig.~\ref{fig:scattering}.
The energy release in the reaction $\psi\psi\to\phi\phi$ is $E_{\rm deep} = 2\Delta$, 
and the corresponding momentum scale is $P_{\rm deep} = (2M \Delta)^{1/2}$.
We are interested in systems consisting of $\psi$~atoms 
whose momenta are small compared to $P_{\rm deep}$.
The reaction $\psi\psi\to\phi\phi$ is therefore a deeply inelastic reaction.
We refer to the momentum scale $P_{\rm deep}$ as {\it high momentum}.

\subsection{Locality}
\label{sec:localdecay}

\begin{figure}[t]
\centerline{ \includegraphics*[width=5cm,clip=true]{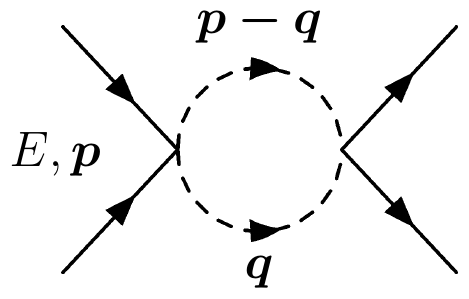} }
\caption{
Diagram of order $g^2$  for the amplitude $i T(E,\bm{p})$ for the transition $\psi\psi \to \psi\psi$
through an intermediate $\phi \phi$ state.}
\label{fig:scattering}
\end{figure}

Because of the large energy release, the deeply inelastic scattering process
$\psi\psi\to\phi\phi$ is effectively local and instantaneous. 
It takes place over a spatial region of size $1/P_{\rm deep}$
and during a time interval of length $1/E_{\rm deep}$.
We proceed to show how this locality can be exploited 
to remove high-momentum $\phi$~atoms from the theory
and construct a low-energy effective field theory for $\psi$~atoms. 
We do this first for two  $\psi$~atoms in this subsection
and then for a system containing many $\psi$~atoms
in subsection~\ref{sec:Effrho}. 
For simplicity, we assume the
coupling~$g$ is small and we work to leading order in~$g$.

The leading contribution to the $\psi\psi\to\psi\psi$
scattering amplitude from a $\phi\phi$ intermediate state  is given by 
the diagram in Fig.~\ref{fig:scattering}.
Using time-ordered perturbation theory
(or using Feynman perturbation theory and integrating by contours over the loop energies),
the off-shell scattering amplitude is 
\begin{equation}
\label{psi-selfenergy}
    T(E,\bm{p}) = - \tfrac{1}{2}
g^2 \int_{\bm{q}} \frac{1}{E - \omega_{\bm{q}} -\omega_{\bm{p}-\bm{q}} + i\epsilon}\,,
\end{equation}
where $E$ and $\bm{p}$ are the total energy and the total momentum 
of the two incoming $\psi$ atoms, respectively.
For convenience, we consider $\psi\psi\to\psi\psi$ scattering
in the center-of-mass frame. In this reference frame, 
the incoming $\psi$ atoms have momenta $\pm\bm{k}$
and energies $E_{\bm{k}}=\bm{k}^2/(2M)$. 
The on-shell scattering amplitude is therefore $T(2E_{\bm{k}},0)$.
The total cross section for $\psi\psi\to\phi\phi$
scattering can be obtained using the optical theorem by evaluating the imaginary part 
of $T$ on the energy shell.
The real part of the scattering amplitude in Eq.~\eqref{psi-selfenergy} is ultraviolet divergent.
The divergence can be canceled by a renormalization of the coupling constant for 
the $\psi\psi\to\phi\phi$ interaction.  After renormalization,
the integral over~$\bm{q}$ in Eq.~\eqref{psi-selfenergy}
is dominated by high momenta of order $P_{\rm deep}$.
Consequently we can expand the on-shell scattering amplitude
$T(2E_{\bm{k}},0)$ in powers of $\bm{k}^2/P_{\rm deep}^2$.
We are primarily interested in constructing an effective Hamiltonian 
that takes into account the leading term $T(0,0)$.
Successively higher powers of $\bm{k}^2/P_{\rm deep}^2$ could 
be taken into account through successively higher-order gradient terms in the effective Hamiltonian.
Similarly, successively higher powers of $E-2E_{\bm{k}}$ in the expansion of $T$
about the on-shell energy could be taken into account through successively higher-order 
time-derivative terms in the effective Lagrangian.

In the sector with only two $\psi$ atoms, the leading effect of the scattering amplitude
$T(E,\bm{p})$ can be taken into account by adding a local term to the free Hamiltonian $H_0^\psi$.
The resulting effective Hamiltonian is
\begin{equation}
\label{eq:heff}
H - i K = H_0^\psi
-  \tfrac14 T(0,0) \int_{\bm{r}}  \psi^{\dagger 2}(\bm{r}) \psi^2(\bm{r}).
\end{equation} 
This equation defines the Hermitian operators $H$ and $K$.
The real part of $T(0,0)$, which is  ultraviolet-divergent,  
can be cancelled by renormalizing
the coupling constant  for the  $ \psi^{\dagger 2} \psi^2$ interaction term
 contained in the ellipses in Eq.~(\ref{eq:Hint}).
The anti-Hermitian part $-iK$ of the  effective Hamiltonian
comes from the imaginary part of $T(0,0)$, which is
\begin{equation}
{\rm Im}  T(0,0) = \frac{Mg^2}{8\pi}(2 M \Delta)^{1/2}.
\label{ImT00}
\end{equation}

The locality of the inelastic scattering process, which  implies $T(2E_{\bm{k}},0)\approx
T(0,0)$,
allows us to simplify the contributions to correlators involving the
operator~$\phi^2(\bm{r},t)$ from its annihilation of $\phi$~atoms that come from $\psi\psi$~scattering.
\begin{figure}[t]
\centerline{ \includegraphics*[width=14cm,clip=true]{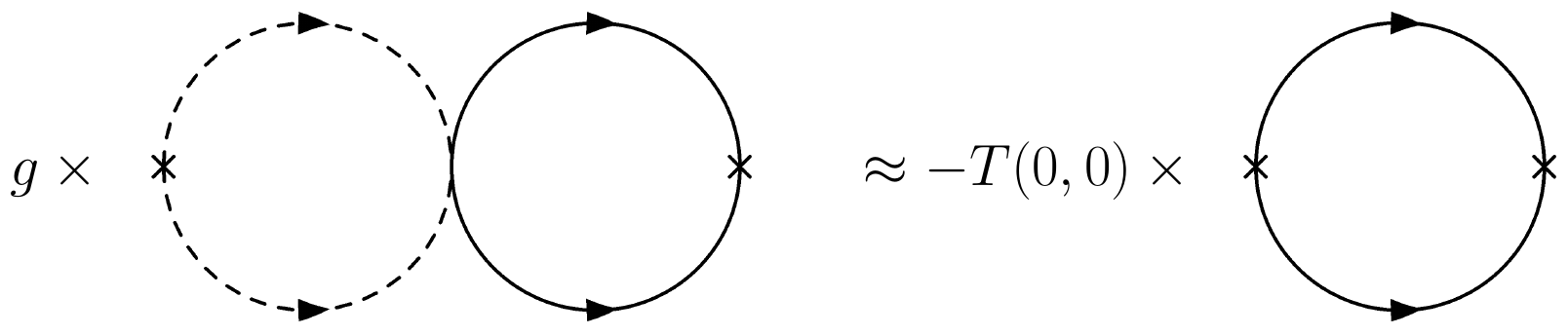} }
\caption{
The correlator $\bra{0}\phi^2(\bm{r},t)\psi^{\dagger2}(\bm{r}',0)\ket{0}$
can be expressed in terms of the 
correlator  $\bra{0}\psi^2(\bm{r},t)\psi^{\dagger2}(\bm{r}',0)\ket{0}$ and the
$T$-matrix for the transition $\psi\psi \to \psi\psi$
through an intermediate $\phi \phi$ state.}
\label{fig:scattering-simp}
\end{figure}
For example, the correlator 
$\bra{0}\phi^2(\bm{r},t)\psi^{\dagger2}(\bm{r}',0)\ket{0}$ illustrated in 
Fig.~\ref{fig:scattering-simp}
can be simplified as follows:
\begin{eqnarray}
\label{eq:phi2psid}
g\,\bra{0}\phi^2(\bm{r},t)\psi^{\dagger2}(\bm{r}',0)\ket{0}
&=&
\int\frac{dE}{2\pi}  \int_{\bm{p}} e^{-iEt+i\bm{p}\cdot(\bm{r}-\bm{r}')}
\int_{\bm{k}} \frac{- 2i \,T(E,\bm{p})}{E - E_{\bm{k}}  - E_{\bm{p}-\bm{k}} + i\epsilon}
\nonumber \\
&\approx&
-T(0,0)\int\frac{dE}{2\pi}\int_{\bm{p}} e^{-iEt+i\bm{p}\cdot(\bm{r}-\bm{r}')}
\int_{\bm{k}} \frac{2i}{E - E_{\bm{k}}  - E_{\bm{p}-\bm{k}} + i\epsilon}
\nonumber \\
&=& -T(0,0)\,\bra{0}\psi^2(\bm{r},t) \psi^{\dagger 2}(\bm{r}',0) \ket{0},
\end{eqnarray}
where $T$ is given in Eq.~(\ref{psi-selfenergy}).
Generally, we can make the following replacements in such situations:
\begin{subequations}
\begin{align}
  g\, \phi^2(\bm{r},t) &\longrightarrow - T(0,0)\,\psi^2(\bm{r},t),
\label{eq:phi^2approx}
\\
g\, \phi^{\dagger2}(\bm{r},t)
&\longrightarrow - T^*(0,0)\,\psi^{\dagger2}(\bm{r},t).
\label{eq:phid^2approx}
\end{align}
\end{subequations}
We use these substitutions repeatedly in the next subsection.

\subsection{Effective Density Matrix}
\label{sec:Effrho}

Replacing the free Hamiltonian $H_0^\psi$ by the effective Hamiltonian $H - iK$ is all that is needed
to analyze the impact of inelastic scattering processes on states with only two $\psi$ atoms.
Analyzing multi-$\psi$ states is more complicated, however, because
a system that is described initially by a state with~$N$ $\psi$~atoms
evolves into a superposition of states with $N$, $N-2$, $N-4$, \ldots\ $\psi$~atoms.
The state with two $\psi$ atoms also evolves into a superposition, but there are only
two states, $N=2$ and~$N=0$, and we do not care about the second one. For
$N>2$, we need the density matrix~$\rho_\full(t)$ to track the superposition of states
containing different numbers of~$\psi$~atoms over time.

A convenient basis for  the quantum-state space of the full theory 
consists of the direct products 
$\ket{\bm{x}_1\ldots \bm{x}_n}_\psi\otimes \ket{\bm{y}_1\ldots \bm{y}_m}_\phi$ of
localized multi-atom $\psi$ and $\phi$ states defined by
\begin{subequations}
\begin{eqnarray}
    \ket{\bm{x}_1\ldots \bm{x}_n}_\psi &=& 
    \frac{1}{\sqrt{n!}}\psid(\bm{x}_n) \cdots \psid(\bm{x}_1) \ket{0}_\psi,
    \label{def-psistates}
    \\
    \ket{\bm{y}_1\ldots \bm{y}_m}_\phi &=& 
    \frac{1}{\sqrt{m!}}\phid(\bm{y}_m) \cdots \phid(\bm{y}_1) \ket{0}_\phi,
\end{eqnarray}
\end{subequations}
where $\ket{0}_\psi$ and $\ket{0}_\phi$ are the vacuum states annihilated by
$\psi(\bm{r})$ and $\phi(\bm{r})$, respectively.
The full density matrix $\rho_\full$ can be expanded in the basis
of direct product states.
Its time evolution is given by
\begin{equation}
\label{rhofull-t}
    \rho_\full(t) \equiv \mathrm{e}^{-iH_\full t}\rho_\full(0)\mathrm{e}^{iH_\full t},
\end{equation}
where $H_\full$ is the full Hamiltonian in Eq.~(\ref{eq:hamiltonian}).
We can define an effective density matrix $\rho(t)$ by tracing over the $\phi$~states:
\begin{subequations}
\begin{eqnarray}
\label{eq:rhoeff}
    \rho(t) 
    &\equiv& \mathrm{Tr}_\phi\,\rho_\full(t)
\\
&=&
 \sum_{m=0}^\infty \int_{\bm{y}_1 \ldots \bm{y}_m}  {}_\phi\bra{\bm{y}_1 \ldots \bm{y}_m}
    \rho_\full(t) \ket{\bm{y}_1 \ldots \bm{y}_m}_\phi.
\end{eqnarray}
\end{subequations}
This operator acts only on $\psi$~states.
A convenient basis for the effective density matrix consists of
outer products of the $\psi$~states defined in Eq.~\eqref{def-psistates}
of the form
$\ket{\bm{x}_1 \ldots \bm{x}_n}_\psi\, {}_\psi 
  \bra{\bm{x}_1^\prime \ldots \bm{x}_{n'}^\prime}$.
The time derivative of the effective density matrix can be obtained by
differentiating Eq.~\eqref{eq:rhoeff} and using the time dependence of the 
full density matrix in Eq.~\eqref{rhofull-t}:
\begin{equation}
\label{eq:ddtrhoeff}
    i\frac{d\ }{dt} \rho = \mathrm{Tr}_\phi\big(
    H_\full\,\rho_\full - \rho_\full\,H_\full \big).
\end{equation}
The contributions
from the kinetic terms in the full Hamiltonian in Eq.~(\ref{eq:hamiltonian}) are simple:
\begin{subequations}
\begin{eqnarray}
\mathrm{Tr}_\phi \big(\big[H_0^\psi, \rho_\full\big] \big)
    &=& \big[H_0^\psi,\rho\big],
\\
\mathrm{Tr}_\phi \big(\big[H_0^\phi, \rho_\full\big] \big)&=& 0.
    \label{Trphi[Hrho]}
\end{eqnarray}
\end{subequations}
The first equation holds because $H_0^\psi$ does not act on $\phi$~states.
The second equation holds because $H_0^\phi$ depends only on $\phi$
fields.\footnote{
The identity $\Tr_\phi ( \hat A \hat B ) = \Tr_\phi ( \hat B \hat A )$
holds for any operator $\hat A$ constructed out of the field $\phi$
and any operator $ \hat B$.  This can be verified by expressing the partial trace 
as a sum over a complete set of $\phi$ states and inserting a complete set of 
$\phi$ states between $\hat A$ and $\hat B$.}
The evolution equation \eqref{eq:ddtrhoeff} reduces to
\begin{equation}
\label{eq:ddtrhoeff-2}
i\frac{d\ }{dt} \rho =  \big[H_0^\psi,\rho\big]
+ \mathrm{Tr}_\phi\big(H_{\rm int}\,\rho_\full - \rho_\full\,H_{\rm int} \big).
\end{equation}

Since there are two terms in the interaction Hamiltonian in Eq.~\eqref{eq:Hint},
there are four contributions to the partial trace in Eq.~\eqref{eq:ddtrhoeff-2}.
We first consider the contribution to the partial trace of 
$H_\mathrm{int}\rho_\full$ from the first term in $H_\mathrm{int}$
in~Eq.~(\ref{eq:Hint}).
This  contribution is dominated by terms in which
$\phi^2(\bm{r})$ annihilates $\phi$~atoms generated (through interactions)
by the $\psi$-sector of~$\rho_\full$, leading to correlators
like~Eq.~(\ref{eq:phi2psid}).
We can therefore use the substitution in Eq.~(\ref{eq:phi^2approx}) to obtain
\begin{eqnarray}
\label{drho/dt-1}
\mathrm{Tr}_\phi\left[\Big(\tfrac{1}{4}g\int_{\bm{r}} \psi^{\dagger 2}(\bm{r})\phi^2(\bm{r})\Big)\rho_\full\right]
 \longrightarrow   -\tfrac{1}{4}T(0,0)\int_{\bm{r}} \psi^{\dagger 2}(\bm{r})\psi^2(\bm{r}) \; \rho.
\end{eqnarray}
Similarly, we can use the substitution in Eq.~(\ref{eq:phid^2approx}) to obtain
the contribution to the partial trace of 
$\rho_\full H_\mathrm{int}$ from the second term in $H_\mathrm{int}$:
\begin{eqnarray}
 \label{drho/dt-2}
\mathrm{Tr}_\phi\left[
\rho_\full \Big(\tfrac{1}{4}g\int_{\bm{r}}\phi^{\dagger 2}(\bm{r})\psi^2(\bm{r})
    \Big) \right]
\longrightarrow -\tfrac{1}{4}T^*(0,0)\int_{\bm{r}} \rho \; \psi^{\dagger 2}(\bm{r})\psi^2(\bm{r}).
\end{eqnarray}
The contributions to the sum of the partial traces in Eqs.~\eqref{drho/dt-1} and \eqref{drho/dt-2}
from the real part of $T(0,0)$ can be added to the term $[H_0^\psi,\rho]$
in Eq.~\eqref{eq:ddtrhoeff-2} to get $[H,\rho]$, where $H$ is the Hermitian part of the 
effective Hamiltonian defined in  Eq.~(\ref{eq:heff}). 
The contributions to the sum of the partial traces in Eqs.~\eqref{drho/dt-1} and \eqref{drho/dt-2}
from the imaginary part of $T(0,0)$ gives $-i\{K, \rho\}$,
where $-iK$ is the anti-Hermitian part of the 
effective Hamiltonian defined in  Eq.~(\ref{eq:heff}). 

Note that we are making a key physical assumption about~$\rho_\full$ when we
use the substitution in Eq.~(\ref{eq:phi^2approx}) to replace 
$g\,\phi^2(\bm{r})$
by $-T(0,0)\,\psi(\bm{r})^2$. As we indicated above, this substitution is valid
provided~$\phi^2(\bm{r})$ annihilates $\phi$~atoms produced by the $\psi$-sector
of~$\rho_\full$.
In principle, it is also possible for~$\phi^2(\bm{r})$ to annihilate
$\phi$~atoms from the $\phi$-sector of~$\rho_\full$. We assume that such contributions
can be ignored because the probability for finding two $\phi$~atoms
at the same space-time point is vanishingly small (and therefore
the probability of an inverse
reaction, $\phi\phi\to\psi\psi$, is negligible).
This is the case, for example, if $\rho_\full$ describes a situation in which all
$\phi$~atoms are produced by inelastic $\psi\psi$ reactions and, once produced,
they either escape from the system
or they interact so weakly with the low-energy $\psi$ atoms that they decouple.

The contribution to the partial trace of $H_\mathrm{int}\rho_\full$ 
from the second term in the interaction Hamiltonian $H_\mathrm{int}$ in Eq.~\eqref{eq:Hint} is
\begin{eqnarray}
\label{drho/dt-3a}
\mathrm{Tr}_\phi\left[\Big(\tfrac{1}{2}g\int_{\bm{r}} \phi^{\dagger 2}(\bm{r})\psi^2(\bm{r})\Big)
\rho_\full\right]
=\mathrm{Tr}_\phi\left[\tfrac{1}{2}g\int_{\bm{r}}\psi^2(\bm{r})\rho_\full  \phi^{\dagger 2}(\bm{r})\right].
\end{eqnarray}
In the second expression, the factor of $\phi^{\dagger 2}(\bm{r})$ has been moved 
to the right side of $\rho_\full$.  To verify this equality,
we begin by inserting the definition of the partial trace on the left side of Eq.~\eqref{drho/dt-3a}:
\begin{eqnarray}
\label{drho/dt-3b}
\sum_{m=0}^\infty \int_{\bm{y}_1 \ldots \bm{y}_m} {}_\phi\bra{\bm{y}_1\ldots \bm{y}_m}
\Big(\tfrac{1}{2}g\int_{\bm{r}} \phi^{\dagger 2}(\bm{r})\psi^2(\bm{r})\Big)
\rho_\full
\ket{\bm{y}_1\ldots \bm{y}_m}_\phi .
\end{eqnarray}
The $\phid(\bm{r})$~factors remove
$\phi$~atoms from the bras on the left side of this equation.
Taking into account the symmetry under permutations of the integration variables, 
their effect can be taken into account by the substitution
\begin{eqnarray}
\label{eq:brasub}
{}_\phi\bra{\bm{y}_1\bm{y}_2\ldots\bm{y}_m} \phi^{\dagger2}(\bm{r})
\longrightarrow  {}_\phi\bra{\bm{y}_1\bm{y}_2\ldots\bm{y}_{m-2}}\,
    \sqrt{m(m-1)} \,\delta^3(\bm{r}-\bm{y}_m)\,\delta^3(\bm{r}-\bm{y}_{m-1}).
\end{eqnarray}
The kets on the right side of Eq.~\eqref{drho/dt-3a} can be expressed as
\begin{eqnarray}
\label{eq:ketsub}
    \ket{\bm{y}_1\bm{y}_2\ldots\bm{y}_m}_\phi =
    \frac{\phid(\bm{y}_m)\phid(\bm{y}_{m-1})}{\sqrt{m(m-1)}}\,\ket{\bm{y}_1\ldots\bm{y}_{m-2}}_\phi.
\end{eqnarray}
Making the substitutions in~Eqs.~(\ref{eq:brasub}) and \eqref{eq:ketsub} and
using the delta functions to integrate over
$\bm{y}_m$ and $\bm{y}_{m-1}$, we obtain the expression on the right side of 
Eq.~\eqref{drho/dt-3a}.  Upon
making the substitution in Eq.~\eqref {eq:phid^2approx},
we obtain
\begin{eqnarray}
\label{drho/dt-3}
\mathrm{Tr}_\phi\left[\Big(\tfrac{1}{4}g\int_{\bm{r}} \phi^{\dagger 2}(\bm{r})\psi^2(\bm{r})\Big)
\rho_\full\right]
\longrightarrow -\tfrac{1}{4}T^*(0,0) \int_{\bm{r}} \psi^2(\bm{r})\,\rho\, \psi^{\dagger 2}(\bm{r}).
\end{eqnarray}
Similarly, the contribution to the partial trace of $\rho_\full H_\mathrm{int}$ 
from the second term in $H_\mathrm{int}$ in~Eq.~(\ref{eq:Hint}) is
\begin{eqnarray}
\label{drho/dt-4}
\mathrm{Tr}_\phi\left[ \rho_\full\Big(\tfrac{1}{4}g\int_{\bm{r}} \psi^{\dagger 2}(\bm{r})\phi^2(\bm{r})\Big)\right]
\longrightarrow -\tfrac{1}{4}T(0,0)\int_{\bm{r}} \psi^2(\bm{r})\,\rho\,\psi^{\dagger 2}(\bm{r}).
\end{eqnarray}
The real part of $T(0,0)$ cancels in the difference between the contributions
in Eqs.~\eqref{drho/dt-3} and \eqref{drho/dt-4}.
The contributions from the imaginary part of $T(0,0)$ 
to the partial trace in Eq.~\eqref{eq:ddtrhoeff-2}
is proportional to the integral over space of the operator $\psi^2(\bm{r})\,\rho\,\psi^{\dagger 2}(\bm{r})$.

Adding the four contributions in Eqs.~\eqref{drho/dt-1}, \eqref{drho/dt-2}, 
\eqref{drho/dt-3}, and \eqref{drho/dt-4}  to the partial trace in Eq.~\eqref{eq:ddtrhoeff-2},
we see that the evolution equation for the effective density matrix is
\begin{eqnarray}
\label{eq:rhoefffinal}
i \frac{d\ }{dt}\rho =  \big[H,\rho\big]
- \frac{i}{4}\,{\rm Im}T(0,0) \int_{\bm{r}} \big[ \psi^{\dagger2}\psi^2(\bm{r})\, \rho 
 + \rho\, \psi^{\dagger2}\psi^2(\bm{r})
-2\, \psi^2(\bm{r})\, \rho\, \psi^{\dagger2}(\bm{r}) \big],
\end{eqnarray}
where ${\rm Im}T(0,0)$ is given in Eq.~\eqref{ImT00}.
This has the standard
form of the Lindblad equation~\eqref{evol:Lindblad}.
The last term removes $\psi$~atoms two at a time
to account for their disappearance due to inelastic scattering into 
pairs of high-momentum $\phi$~atoms.

\end{document}